\documentclass{Hou-ws-p10x7}

\def\ltap{\ \raisebox{-.5ex}{\rlap{$\sim$}} \raisebox{.4ex}{$<$}\ }
\def\gtap{\ \raisebox{-.5ex}{\rlap{$\sim$}} \raisebox{.4ex}{$>$}\ }

\def\prd#1#2#3{    {\it Phys. Rev. }{\bf D #1}, #3 (#2)}

\def\prl#1#2#3{    {\it Phys. Rev. Lett. }{\bf #1}, #3 (#2)}

\begin{document}

\title{POSSIBILITY OF LARGE FSI PHASES IN LIGHT OF 
	{\boldmath $B\to K\pi$ \& $\pi\pi$}
	 DATA}

\author{GEORGE W.S. HOU}

\address{Department of Physics, National Taiwan University,
Taipei, Taiwan, R.O.C. 
}

\twocolumn[\maketitle\abstract{
After briefly reviewing how data indicated that 
factorization seems to work in observed two body charmless modes, 
{\it if one takes $\gamma \gtap 90^\circ$},
we point out that the $K^0\pi^0$ mode seems too large.
This and other hints suggest that perhaps not only $\gamma$ is large,
but rescattering phase $\delta$ could be sizable as well.
}]

\section{Path to $\gamma > 90^\circ$ and Factorization
}

CLEO data has driven B phenomenology in the classic way in 
the last 3 years.

\vskip0.1cm
\noindent{\bf \underline{1997}}: $\bar K^0\pi^- > K^-\pi^+
				   \simeq 1.5\times 10^{-5}$\\
This lead to the Fleischer--Mannel bound
and a boom in theory activity,
leading eventually to model-indep. methods 
for extracting $\gamma$.

\vskip0.1cm
\noindent{\bf \underline{1998}}: 
$\bar K^0\pi^- \simeq K^-\pi^+ \ltap K^-\pi^0 \simeq 1.5\times 10^{-5}$\\
First equality precipitated suggestion for large $\gamma$;
strength of $K\pi^0$ indicated EWP.\cite{DHHP}

\vskip0.1cm
\noindent{\bf \underline{1999}}: Multiple modes emerge, {\it e.g.}\\
$\star$\ $\rho^0\pi$, $\rho\pi$, $\omega\pi$:\ \
	   $\exists\ b\to u$ tree (T).\\
$\star$\ $\omega K$ disappear:\ \ As it should.\\
$\star$\ $K\pi^0\simeq {2\over 3}\, (\bar K^0\pi\simeq K\pi)$:\ \ 
  EWP at work!\\
$\star$\ $\pi\pi \sim {1\over 4}\, K\pi$
  $\Longrightarrow$ Large $\gamma$!\\
$\star$\ $K^0\pi^0 \sim K\pi,\ K^0\pi \Longrightarrow$
	 \underline{Problem}.

The host of emerging modes in 1999 lead to 
the observation\cite{HHY} that
``{\it Factorization\ works\ in\ observed\
two\ body\ charmless\ rare\ B\ decays,\
{\bf if}\ $\cos\gamma \ltap 0$}."
To stay low key,
only the sign change in $\cos\gamma$ was 
initially advocated,
but emboldened by this observation,
quantification was sought. 
Using known factorization formulas {\it etc.},\cite{Ali}
a ``global fit" of more than 10 modes
gave\cite{HSW} $\gamma \simeq 105^\circ$,
which is in some conflict with the ``CKM Fit" value\cite{Stocchi} 
of $\gamma\simeq 58.5^\circ\pm 7.1^\circ$.
However, by end of 1999,
all B practioners have switched to 
$\gamma \gtap 80^\circ$--$90^\circ$,
as reflected in the 5 rare B theory talks here.

We have gained from {\it hadronic} rare B modes 
new knowledge on CKM.
That hadronization does not mask this (factorization works!)
is quite astonishing.
At this Conference, first physics results have been reported 
from (asymmetric) B Factories. 
Both BaBar and Belle have collected data 
comparable to CLEO II$+$ II.V.
Belle\cite{Paoti} confirms CLEO results on $K\pi$ modes and $\pi^-\pi^+$,
while BaBar\cite{Champion} is at some variance, finding
$\pi^-\pi^+ \simeq K^-\pi^+$.
One surprise is the $\phi K$ mode,
where Belle reports a sizable rate with {\it lower bound} 
above the new CLEO central value,\cite{Stroy}
which is just above its own previous limit.
The age of competition has obviously arrived, 
and we look forward to healthy and 
at same time explosive developments in near future.

It is said\cite{Cheng} that
$\rho^0\pi \ltap \omega\pi$ at present
no longer supports $\gamma > 90^\circ$
as it was\cite{HHY} in early 1999 when
$\rho^0\pi > \omega\pi$ was reported by CLEO.
Likewise, $\eta^\prime K^0 > \eta' K^-$ is also
at odds with $\gamma > 90^\circ$.
We caution that the number of $\omega\pi$ events 
reported by BaBar\cite{Champion}
indicates a rate lower than CLEO's, while 
the $\rho\pi$ rates seem considerably larger.
As for the $\eta^\prime K$ modes,
they are not yet fully understood.
As stressed by Golutvin,\cite{Gol}
clearly we ``NEED MORE DATA!", which will arrive in due course.

\section{Problems?}
\vskip0.1cm

Besides the strength of $\eta^\prime K$ modes,
another problem was apparent by summer 1999:
$K^0\pi^0$ seems too large!\cite{HY}
Playing games
\footnote{
We are well aware of the ``Central Value Syndrome"
sufferred by theorists,
but advocate that these games still stimulate the field
by exploring possibilities.
}
with ``central values" from CLEO,
we might also note that $\pi^-\pi^+ < \pi^-\pi^0$ seems a bit small,
while the direct CP asymmetries $a_{\rm CP}$
in $K^-\pi^+$, $K^-\pi^0$ and $\bar K^0\pi^-$ modes
give a ``pattern" that is different from
SM expectations with only S.D. rescattering phases.

As mentioned, $K\pi^0/K\pi \simeq 0.65$
confirms {\it constructive}
EWP-P interference for $K\pi^0$ in SM.
From the operators and the $\pi^0$ w.f. 
(sign traced to $d\bar d\to \pi^0$)
one expects {\it destructive} EWP-P interference
in $K^0\pi^0$,\cite{HY}
\begin{eqnarray}
\frac{\overline K^0\pi^0}{ \overline K^0\pi^-}
 \approx {1\over 2}
\left\vert 1 - r_0 \,
 {
   1.5 a_9
\over   a_4 +a_6 R}\right\vert^2
 \approx  {1\over 3},
\end{eqnarray}
where $r_0 = f_\pi F_0^{BK}/ f_K F_0^{B\pi} \simeq 0.9$ and
$R = 2{m_K^2/(m_b-m_d)(m_s+m_d)}$.
Hence $K^0\pi^0 > K\pi^0$
is very hard to reconcile.

Could this be due to Final State Interactions (FSI) alone?
Taking $\gamma = 60^\circ$,
$\pi^-\pi^+$ can be accounted for
via $\pi^-\pi^+ \leftrightarrow \pi^0\pi^0$ rescattering,
but the $K\pi$ modes fit data poorly.
Thus, we still need to call on the service of $\gamma$.

\section{Large {\boldmath $\gamma$} and {\boldmath $\delta$?}}

Let us set up a simple formalism for what we mean by FSI,
or $\delta \neq 0$, which goes beyond factorization.
Our {\it Ansatz} simply extends
naive factorization amplitudes $A_I$ by
adding $\delta_I$ to model hadronic phases in final state.
\footnote{
S.D. quark-level rescatterings 
are included in $A_I$.
}
The $B\to \pi\pi$ amplitudes become,
\begin{eqnarray}
 {\cal A}(B\to \pi\pi)
 &=&A_0e^{i\delta_0}
  + A_2 e^{i\delta_2},\nonumber\\
 {\cal A}(B\to \pi^0\pi^0)
 &=&\mbox{\small $1\over \sqrt{2}$}A_0e^{i\delta_0}
           -\mbox{\small $\sqrt{2}$}A_2 e^{i\delta_2},\nonumber\\
 {\cal A}(B\to \pi\pi^0)
 &=&\mbox{\small $3\over \sqrt{2}$}A_2 e^{i\delta_2},
 \end{eqnarray}
where $I = 0,\ 2$ stand for final state isospin.
For $K\pi$ modes, we have the amplitudes
\begin{eqnarray}
 {\cal A}(B\to K\pi)
 &=&A_{3\over 2} e^{i\delta_{3\over 2}}
  - A_{1\over 2}^- e^{i\delta_{1\over 2}},
  \nonumber\\
 {\cal A}(B\to K^0\pi^0)
 &=&\mbox{\small $\sqrt{2}$}A_{3\over 2} e^{i\delta_{3\over 2}}
  + \mbox{\small $1\over \sqrt{2}$}A_{1\over 2}^- e^{i\delta_{1\over 2}},
   \nonumber\\
  {\cal A}(B\to K\pi^0)
  &=&\mbox{\small $\sqrt{2}$}A_{3\over 2} e^{i\delta_{3\over 2}}
   + \mbox{\small $1\over \sqrt{2}$}A_{1\over 2}^+ e^{i\delta_{1\over 2}},
  \nonumber\\
 {\cal A}(B\to K^0\pi)
 &=&-A_{3\over 2} e^{i\delta_{3\over 2}}
   + A_{1\over 2}^+ e^{i\delta_{1\over 2}},
\end{eqnarray}
where $A_{1\over 2}^\mp \equiv A_{1\over 2} \mp B_{1\over 2}$
and $A_I\, (B_I)$ are
$\Delta I =1\ (0)$ amplitudes for final state isospin $I$.
It is tempting to use SU(3), since
$\delta_{2}\cong \delta_{3\over 2}$ holds.
However, $(\pi\pi)_{I=0}$ has {\bf 1} in addition
to {\bf 8} and {\bf 27} contributions, 
so in principle $\delta_{0}\neq \delta_{1\over 2}$.
Furthermore, $K\bar K$ has yet to be seen.

\begin{figure}[t!]
\centerline{
            {\epsfxsize2.7 in \epsffile{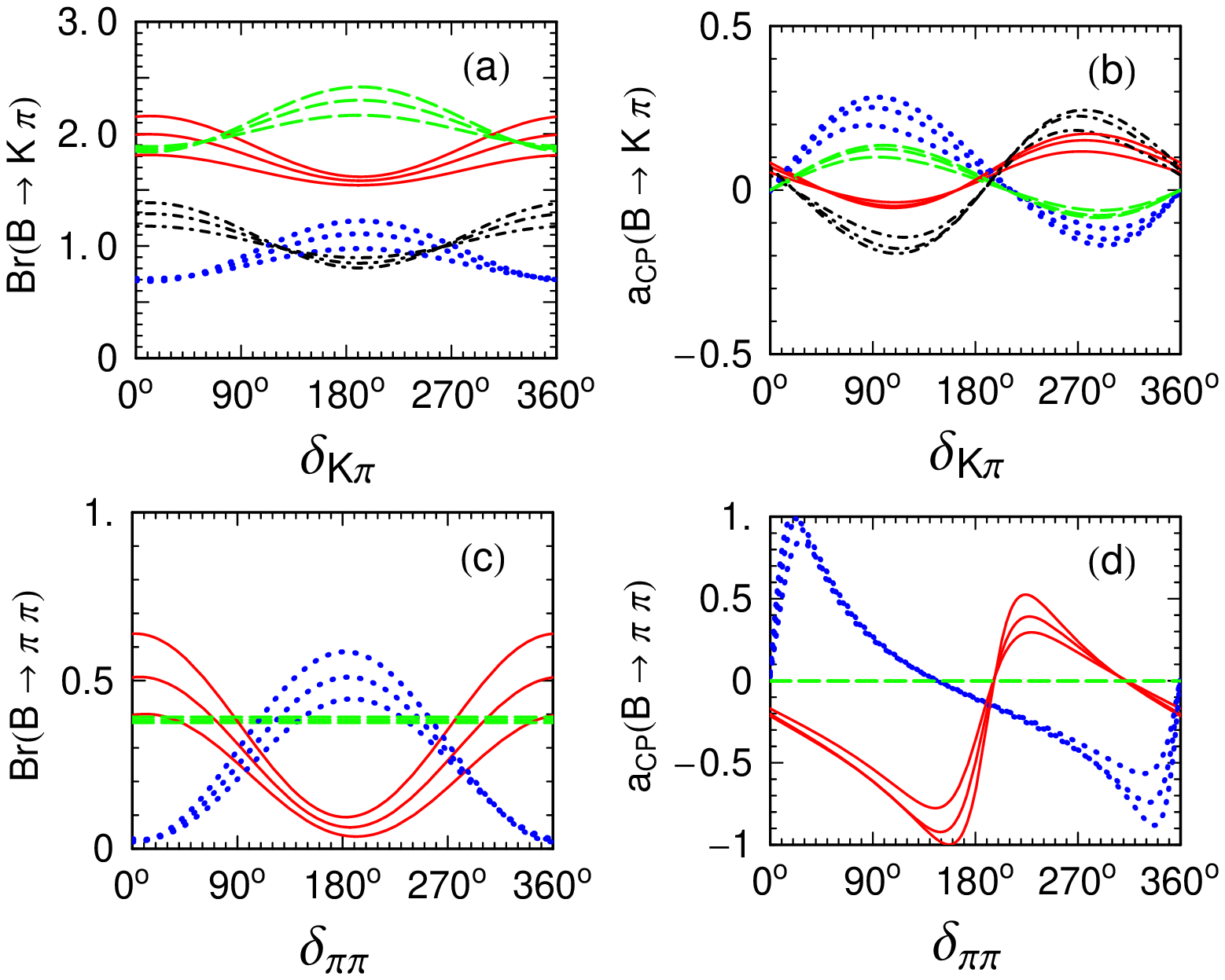}}}
\smallskip
\caption {Brs and $a_{\rm CP}$s for $K\pi$ and $\pi\pi$ vs.
$\delta$.
For curves from:
 (a) up (down) to down (up) for
$K^-\pi^+, \overline K^0\pi^{-,0}$ ($K^-\pi^0$) at
$\delta=180^\circ$; and
 (b) up (down) to down
(up) for $K^-\pi^{+,0}$
    ($\overline K^0\pi^{-,0}$) at
$\delta=90^\circ$; and
 (c) down to up at $\delta=180^\circ$; and
 (d) down to up for $\pi^-\pi^+$ ($\pi^0\pi^0$)
     at $\delta=160^\circ$ ($20^\circ$),
are for
$\gamma=130^\circ, 110^\circ$, and $90^\circ$, respectively.}
\label{fig:gammalarge}
\end{figure}

We plot in Fig.~\ref{fig:gammalarge} the Brs and
$a_{\rm CP}$s vs. the phase differences
$\delta_{K\pi}=\delta_{3\over 2}-\delta_{1\over 2}$ and
$\delta_{\pi\pi}=\delta_2-\delta_0$, respectively, 
for several large $\gamma$ values.
A numerical exercise illustrates the point.
With CLEO data only 
(Belle/BaBar results are still too preliminary),
one has $K\pi : K^0\pi : K\pi^0 : K^0\pi^0 =
1 : 1.06 : 0.67 : {\bf 0.85}$.
Taking, {\it e.g.} $\gamma = 110^\circ$,
we read off from Fig. 1(a) and 
find the ratio $1 : 0.94 : 0.65 : 0.35$,
and $K^0\pi^0$ is clearly a problem. 
Allowing $\delta_{K\pi}\sim 90^\circ$, 
the ratio becomes $1 : 1.12 : 0.61 : 0.47$.
This is far from resolving the problem, 
but the {\it trend} is good.
What's more, the central values for $a_{\rm CP}$ in
$K\pi$, $K^0\pi$, $K\pi^0$ modes become
$-0.04$, $0.13$ and $-0.16$, which fits
the pattern of experimental central values
$-0.04$, $0.17$ and $-0.29$ very well. 
We see from Fig. 1(c) that taking 
$\delta_{\pi\pi} \sim \delta_{K\pi}$
gives $\pi\pi < \pi\pi^0$.
There are some further dramatic consequences:

\noindent \phantom{i} $\bullet$
 $a_{\rm CP}^{\bar K^0\pi^0} \sim - a_{\rm CP}^{K^-\pi^0}$ large.

\noindent \phantom{i} $\bullet$
 $\pi^0\pi^0 \sim \pi\pi \sim 3$--$5 \times 10^{-6}
	 \ltap \pi\pi^0$.
	 
\noindent \hskip1.2cm {\small (still satisfy CLEO bound)}

\noindent \phantom{i} $\bullet$
 $a_{\rm CP}^{\pi\pi},\ a_{\rm CP}^{\pi^0\pi^0}
	 \sim -60\%$, $-30\%$ possible.\\
{\it measurable in a couple of years}.

\section{Remarks and Comments}

Our approach of elastic $2 \to 2$ rescattering 
may be too simplistic, since
$B\to i \to K\pi$ involves many intermediate staes $i$
and can be highly inelastic.
\footnote{
``Pomeron" exchange suggest 
$\pi/2$ phase shift for all channels
hence $\delta \simeq 0$,
while ``Regge" exchange is subleading,
typically giving $\delta \sim 10^\circ$--$20^\circ$.
}
Furthermore,
our elastic rescattering is of form
$K^-\pi^+\to \bar K^0\pi^0$.
Such ``charge exchange" for $p_{K,\pi} \sim 2.5$ GeV is rather
counter-intuitive.
We stress, however,
that our approach is {\it phenomenological}:
data indicates large $\gamma$ plus simple factorization works; 
we then make a minimal extension with FSI phases,
without pretending to know their origin.
They could be effective parameters arising from 
e.g. annihilation diagrams.\cite{Hn}
But if they genuinely arise from L.D. physics,
they would then pose a real problem for PQCD,
which argues that long distance effects
are $1/m_b$ suppressed.\cite{Beneke}

We offer some brief comments:

\noindent \phantom{i} $\bullet$
 Some people define $\delta = \delta_{\rm P} - \delta_{\rm T}$.
 For us this mixes elastic and inelastic phases.

\noindent \phantom{i} $\bullet$
Many works {\it force} $K^0\pi$ to be pure P,
which is a strong assumption {\it vs.} Eqs. (2) and (3).

\noindent \phantom{i} $\bullet$
To account for large $K^0\pi^0$, 
some recent work\cite{ZWNG} 
invoke a ``$a_{3\over 2}^c$" amplitude that is
8 to 10 times larger than factorization result
(which arises from the isospin violating EWP),
where $c$ indicates it arises from charm intermediate states.
We point out, however, that
$b\to c\bar cs$ is purely $\Delta I = 0$ and cannot generate
$I = {3\over 2}$ final state.

\noindent \phantom{i} $\bullet$
$D\bar D\to \pi\pi$ rescattering has been used\cite{Xing}
to illustrate the importance of inelastic rescattering.
While it is fine as an illustration,
a single channel is not quite meaningful
because of existence of many channels.

We believe that inelastic phases are {\it impossible}
to understand precisely because there are too many channels.
A statistical approach\cite{SuzWolf} of 
averaging over large number of random inelastic phases 
again gives $\delta \sim 10^\circ$--$20^\circ$.
In contrast,
$2\to 2$ elastic rescattering is unique for two-body final states.
To the least it effectively models hadronic interactions
beyond (naive) factorization,
which seems to work in two body charmless B decays so far.

\section{Conclusion}

Data indicates that factorization
seems to work for the first 10-20 or so two body rare B modes,
{\it if} we take $\gamma \gtap 90^\circ$.
An exception is the strength of $K^0\pi^0$ mode.
We find a coherent picture where
$\gamma$ is large, but so is some effective FSI phase $\delta$.
The picture can account for the
current central values of $\pi^-\pi^+$ vs. $\pi^-\pi^0$
and the pattern seen in $a_{\rm CP}$'s for $K\pi$ modes.
It also gives consequences that are testable in next couple of years:
Large $a_{\rm CP}$ in 
$\bar K^0\pi^0$ and $K^-\pi^0$;
$\pi^0\pi^0 \sim \pi^-\pi^+$
with $a_{\rm CP}$ as large as $-30\%$ and $-60\%$, respectively.

\end{document}